\documentstyle[epsfig]{mn}

\loadboldmathitalic
\title[Cosmological Magnetic Pressure]{The Growth of Baryonic Structure in the
presence of Cosmological Magnetic Pressure}
\author[Gazzola \etal]{Lorena Gazzola\thanks{E-mail:ppxlg\@@nottingham.ac.uk}, Emma J. King, Frazer R. Pearce \&
Peter Coles \\ School of Physics \& Astronomy, University of
Nottingham, University Park, Nottingham, NG7 2RD}

\date{\today}
\def\lesssim{\mathrel{\hbox{\rlap{\hbox{\lower4pt\hbox{$\sim$}}}\hbox{$<$}}}}
\def\gtrsim{\mathrel{\hbox{\rlap{\hbox{\lower4pt\hbox{$\sim$}}}\hbox{$>$}}}}
\def\ion#1#2{#1$\;${\small\rm\@@roman{#2}}\relax}
\def\etal{{\it et al.\thinspace}}

\begin{document}
\maketitle
\begin{abstract}\\
We follow the growth of baryonic structure in the presence of a
magnetic field within an approximate cosmological
magneto-hydrodynamic simulation, produced by adding an (isotropic)
magnetic pressure related to the local gas pressure. We perform an ensemble of
these simulations to follow the amplification of the  field with
time. By using a variety of initial field strengths and changing the
slope of the power law that governs the way the field grows with
increasing density we span the range of current observations and
demonstrate the size of the effect realistic magnetic fields could
have on the central density of groups and clusters. A strong
magnetic field significantly reduces the central gas density which,
in turn, reduces observable quantities such as the X-ray luminosity.
\end{abstract}


\section{Introduction}\large

In an era where we have such a firm grasp of many of the fundamental
parameters that define the universe in which we live (Spergel \etal
2003), the underlying framework of cosmology based on gravitational
instability of cold dark matter is no longer subject to serious
doubt. However, although the idea that large potential wells are
built up by the hierarchical assembly of many thousands of smaller
sub-blocks (Press \& Schechter 1974; Davis \etal 1985) has not been
seriously challenged for decades, there is much current debate about
the presence or absence of large amounts of substructure
(Kazantzidis \etal 2004) and about the cuspiness of the inner parts
of dark matter profiles (Moore \etal 1999; Power \etal 2003;
Diemand, Moore \& Stadel 2004). Moreover, the reaction of baryonic
material to this underlying framework is not well understood at all.
We still do not readily understand why such a small fraction of
baryons cools to form stars (Fukugita, Hogan \& Peebles 1998; Balogh
\etal 2001), how the first stars are distributed (Abel, Bryan \&
Norman 2002) and subsequently reionize the universe (Sokasian \etal
2004) or why the gas at the centre of dark matter haloes obstinately
refuses to cool efficiently (Ponman \etal 1999; Pearce \etal 2000,
2001; Voit \& Bryan 2001). 

An additional complication arises from the well-known fact that the
intra-cluster medium (ICM) in galaxy clusters is magnetized. This is
inferred from observations of diffuse radio haloes and hard X-ray
emission as well as Faraday rotation measurements (Kronberg 1994;
Feretti \& Giovannini 1996; Govoni \etal 2001a; Taylor \etal 2001;
Eilek \& Owen 2002; Murgia \etal 2004; Fusco-Femiano \etal 2004).
Magnetic fields are observed throughout the Universe at a wide range
of epochs and scales; from a few $\mu G$ up to a few $mG$ within
galaxies (Beck 2000; Krause 2003), and from a few tenths of a $\mu
G$ on cluster and super-cluster scales, to values in the centre of
cooling flow clusters reaching as high as a few tens of $\mu G$
(Carlstrom, Holder \& Reese 2002). At high redshifts magnetic fields
of a few $\mu G$ have been observed in systems such as
Lyman-$\alpha$ absorption systems, (Kronberg \& Perry 1982; Norman
1990; Kronberg 1994; Oren \& Wolfe 1995), radio-galaxies (Athreya
\etal 1998; Pentericci \etal 2000), and proto-clusters (Pentericci
\etal 2000; Bagchi \etal 2002).  Although magnetic fields of the
order of $\mu G$ in strength may seem insignificant, the vast scales
involved mean that a significant amount of energy is stored in these
fields, and consequently they can be dynamically important. Where
they come from is still a mystery. In this paper we shall assume
that some primordial seed field exists and experimentally determine
whether it can have a significant effect on cluster properties.

Analytical and numerical studies provide an important counterpoint
to these observations, aiding our understanding of the distribution and
perhaps origin of the magnetic field. The first attempt to implement
MHD (magneto-hydrodynamics) into SPH (smoothed particle
hydrodynamics) was the polytropes studied by Gingold \& Monaghan
(1977). Further aspects of the application of SPH to magnetic
phenomena were considered by Phillips \& Monaghan (1985), Phillips
(1986) and Monaghan (1992).  Phillips \& Monaghan demonstrated that
when the MHD equations are written in a conservative form a
numerical effect occurs: SPH particles tend to clump due to the
presence of an artificial tension. Subsequent authors (Meglicki,
Wickramasinghe \& Dewar 1995; Cerqueira \& de Gouveia Dal Pino 2001;
Hosking 2002) circumvented this problem using non-conservative
forces but this approach does not work well when shocks are present.
More recently, Price \& Monaghan (2004) developed a new approach in
which they added a small artificial stress that prevented the
numerical instability.

Detailed MHD simulations of the growth of structure in the universe
pointed out the importance of merger events on the distribution and
strength of the final magnetic field and that the initial field
structure is completely wiped out during cluster formation. In
addition, compression and shear flows can strongly amplify the seed
field (Birk, Wiechen \& Otto 1999; Roettiger, Stone \& Burns 1999;
Dolag, Bartelmann \& Lesch 1999, 2002).

Loeb \& Mao (1994) have suggested that, if the resultant magnetic
field were sufficiently strong, the ICM could be supported to a
significant extent by magnetic pressure in addition to thermal gas
pressure. In clusters where this was the case the magnetic field
would be dynamically important (Dolag \etal 2001; Eilek \& Owen
2002), so this support might contribute to the discrepancy between
X-ray and gravitational lensing mass estimates of the central
regions of galaxy clusters. Gon\c calves \& Fria\c ca (1999) found
that the magnetic field could be dynamically important on scales as
small as $\leq 1 \;{\rm kpc}$, although it seems unlikely that
either these small scale fields or a more widely distributed field
such as that proposed by Loeb \& Mao could be the main reason for
the discrepancy between the mass estimates in the central regions,
at least in relaxed clusters (Dolag \etal 1999; Dolag \& Schindler
2000).

In this paper we examine the effect of introducing a large-scale
magnetic field on the growth of baryonic structure. As gas collapses
isotropically the magnetic field strength grows as $B \propto
\rho^{2 \over 3}$ assuming that the field lines are frozen into the
plasma. Accordingly, since the magnetic pressure is $P_{\rm mag}
\propto B^2$, the additional gas pressure due to the presence of a
magnetic field rises as $\rho^{4 \over 3}$. Here we implement this
simple assumption into an $N-$body hydrodynamics code and follow
what happens to the density of each of the gas particles. Obviously
this method relies on a smooth collapse of the gas without massive
tangling of the magnetic field lines, a situation that restricts its
applicability to relatively low values of the overdensity. For this
reason we have made no attempt to follow gas cooling or galaxy
formation, but have concentrated on the hot haloes of galaxy groups
and clusters. In order to mimic more realistic fields we parametrise
our model so that the power-law index for the $B-\rho$ relation is
one of the free parameters, $\alpha$; the motivation for this is
discussed in the next section, where we detail our numerical models
and magnetic field approximation.  Our results are presented in
Sections 3 \& 4, where we study various initial field strengths and
power law dependencies with density as well as varying the redshift
at which the field is initiated. Section 5 contains a discussion and
our conclusions.

\section{Numerical Calculations}

\subsection{The Simulation Method}
The basis for our model of the passive evolution of magnetic fields
is Hydra, an $AP^{3}M-SPH$ code (Couchman, Thomas \& Pearce 1995). Smoothed-particle
Hydrodynamics (SPH) is a Lagrangian numerical method which follows
the motion of a set of fluid (gas) elements represented by discrete
particles. The thermal energy and velocity of each particle are
known at any given time and each particle has a fixed mass.
Properties of the gas at the position of a particle can be estimated
by smoothing these quantities over the $N_{\rm{SPH}}$ nearest
neighbouring particles. The gas properties are then used to
calculate the forces acting on each particle in order to update the
positions and velocities. In cosmological simulations both dark
matter and gas particles are included and the particles are
initially distributed in a manner consistent with a cosmological
power spectrum. If the process of galaxy formation is to be
simulated then radiative cooling of the gas must also be included;
we neglect this for the purposes of this paper.

\begin{figure}
\begin{center}
\leavevmode \epsfysize=8cm \epsfbox{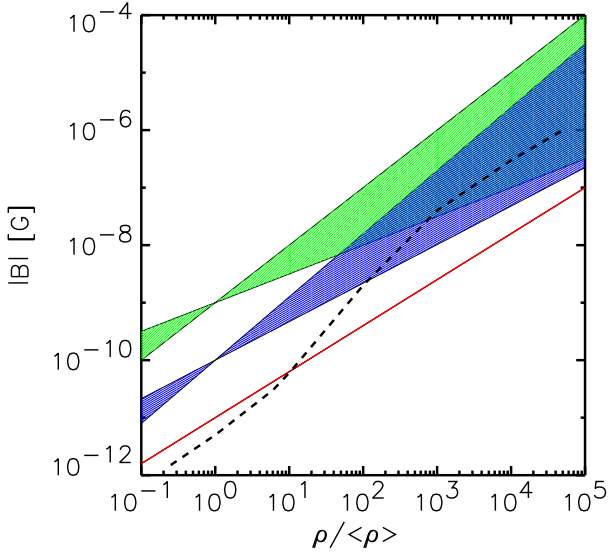}
\end{center}
\caption{Magnetic field strength as a function of overdensity. The
shaded regions and the continuous line delimit our range of $\alpha$ and $B_{0}$. Top green: $B_{0}=10^{-9} G$ and $0.5 \le \alpha \le 1.$; bottom blue: $B_{0} =10^{-10} G$ and $2/3 \le \alpha \le 1.1$; red continuous line: $B_{0}=10^{-11} G$ $\alpha=0.8$. The black dashed line is from Dolag et al. 2005.
\label{figmagprop}}
\end{figure}

The base simulation used throughout this paper has $128^3$ gas and
$128^3$ dark matter particles with individual masses of $6.58 \times
10^{8}h^{-1} {\rm M_{\odot}}$ and $4.27 \times 10^{9} h^{-1} {\rm
M_{\odot}}$ respectively, in a periodic box of $50h^{-1}{\rm Mpc}$.
The power spectrum is that appropriate to a cold dark matter
universe with the following parameter values: mean mass density
parameter $\Omega_M=0.3$, cosmological constant
$\Omega_{\Lambda}=0.7$, baryon density parameter $\Omega_b=0.04$,
Hubble constant (in units of $100 {\rm kms^{-1}Mpc^{-1}}$), $h=0.7$,
power spectrum shape parameter $\Gamma=0.21$ and rms linear
fluctuation amplitude $\sigma_8=0.90$. These cosmological parameters
are close to those obtained through fits to {\it WMAP} data (Spergel
\etal 2003): $\Omega_M=0.27$, $\Omega_{\Lambda}=0.73$,
$\Omega_b=0.044$, $h=0.71$, $\sigma_8=0.84$. The gravitational
softening length is $20 h^{-1}\rm{kpc}$, fixed in physical
coordinates. The starting redshift was $z_{\rm start}=49$.

We carried out a set of $9$ runs, each one characterised by different parameter values (see Table ~\ref{tab1} and following section). For each run we extracted clusters with a spherical overdensity criteria and selected those with at least $N=1000$ particles of each kind within the virial radius. The number of objects we analysed is of the order of $140$ per run.    Given the limited volume of our simulations, we have only a couple of moderately rich objects, with ${\rm M_{vir}} > 10^{14}h^{-1} {\rm M}_{\odot}$.
\begin{table}
\begin{tabular}{lcc}
\hline
{\bf Model}       &{\bf Field normalisation}  &{\bf slope } \\
            & $B_{0}$ [Gauss] &$\alpha$  \\
\hline
noB         &0            &-\\
\\
B0960       &$10^{-9}$    &1.00\\
B0948       &$10^{-9}$    &0.80\\
B0940       &$10^{-9}$    &2/3\\
B0930       &$10^{-9}$    &0.50\\
\\
B1066       &$10^{-10}$   &1.10\\
B1060       &$10^{-10}$   &1.00\\
B1048       &$10^{-10}$   &0.80\\
B1040       &$10^{-10}$   &2/3\\
\\
B1148       &$10^{-11}$   &0.80\\

\hline
\end{tabular}
\caption{Parameters for the simulations used in this paper. By column
a simulation identifier, the magnetic field normalization in Gauss ($B_0$)
and the power law index of the field$-$density relation ($\alpha$).
\label{tab1}}

\end{table}

\subsection{Approximating a Magnetic field}
In this section we describe how we introduce a magnetic field into
our simulations and justify our choice of parameters.

According to the standard picture of magnetic field evolution a seed
field generated at high redshift is amplified by adiabatic
compression and merger events occurring during cluster formation. A
straightforward analytical argument shows that under the assumptions
of:
\begin{itemize}
\item{B-field frozen into plasma} \item{uniform spherical
collapse} \item{small magnetic field}
\item{mass conservation}
\item{magnetic flux conservation},
\end{itemize}
then $B$ scales with the gas density $\rho$ as $B \propto
\rho^{\alpha}$ with $\alpha=2/3$. On the other hand, if the magnetic
field is important, there will be a preferential axis $x$ for the
collapse due to the presence of strong field lines and the collapse
proceeds {\it cylindrically}. This, together with flux conservation
and the assumption that gravity is in equilibrium with thermal
pressure along the symmetry axis ($2\pi G\rho x^{2} \sim c_{s}^{2}$)
leads to $B \propto (T\rho)^{1 \over 2}$, so if the object is
isothermal the power-law relation between $B$ and $\rho$ acquires a
different value of $\alpha = 1/2$ (Crutcher 1999). 

More detailed numerical studies have found that in realistic large-scale
structures without special geometries the amplification of the seed
field  is not quite as expected from these simple collapse models,
with a rough power-law dependence but a higher value of $\alpha$
(Roettiger \etal 1999; Dolag \etal 1999; Dolag \etal 2005). King \&
Coles (2006) showed that an average over small-scale collapses leads
to an average value of $\alpha$ which is higher than the isotropic
value $\alpha=2/3$; the average of a non-linear function is not the
same as the non-linear function of the average. Moreover, other
phenomena such as mergers and anisotropies act to enhance the field
still further. It is therefore likely that different power indices
better describe the field behaviour in different density ranges, as
found observationally: Dolag \etal (2001) found $\alpha = 0.9$ for
the galaxy cluster Abell 119, while Crutcher (1999) got $\alpha =
0.47$ for molecular clouds in a density regime of about
$10^{3}-10^{4} {\rm cm^{-3}}$. 

The parametrisation we have chosen
corresponds to a polytropic equation of state of the form $P\propto
\rho^{2\alpha}$; large values of $\alpha$ will therefore produce a
steeper relation than pertains for the thermal pressure ($\alpha
\simeq 5/6$). This is the regime in which interesting physics can
occur.

These considerations all lead us to approximate the magnetic field
$B(z)$ as:
\begin{equation}
B(z)=B_{0}(1+z)^{3\alpha}\rho^{\alpha} ,
\label{eqfield}\end{equation}
where $\rho$ is the local overdensity, $B_{0}$ the field in an
unperturbed universe, both in comoving units, and $z$ is the
redshift. The term $\rho^{\alpha}$ accounts for the field
amplification by incorporating $\alpha$ as a free parameter. By
varying the power-law index we emulate the different possible
amplification mechanisms, though we stress that we do not solve the
full MHD equations exactly, in particular we ignore the back-reaction of the field on the gas which would break the simple scaling of equation (\ref{eqfield}). In reality it is likely that the
amplification of a $B$-field by the factors we have discussed would
saturate around the equipartition value. Care must be taken,
therefore, in interpreting our results when magnetic pressure
dominates the thermal pressure. We are primarily interested,
however, in the growth of structures through the linear and
quasi-linear regime so we hope our calculations have a reasonable
regime of validity. It is also important to understand that we can
not use this method to simulate ordered fields displaying
large-scale anisotropy: we are restricted to fields that are tangled
on a sufficiently small scale that their effect on the large-scale
gas motion is isotropic.

The standard SPH momentum equation for particle $i$ using a
smoothing kernel $W$ is:
\begin{equation}
\frac{d\vec{\nu_{i}}}{dt} = -\sum_{j} m_{j}
\left(\frac{P_{i}}{\rho_{i}^{2}} +\frac{P_{j}}{\rho_{j}^{2}} +
\prod_{ij} \right) \nabla_{i}W_{ij} ,
\end{equation}
and the corresponding equation for the rate of change of the
internal energy $e_{i}$ is
\begin{equation}
\frac{de_{i}}{dt} = \frac{1}{2}\sum_{j} m_{j}
\left(\frac{P_{i}}{\rho_{i}^{2}} +\frac{P_{j}}{\rho_{j}^{2}}
 + \prod_{ij}\right) \vec{\nu_{ij}}\cdot\nabla_{i}W_{ij},
\end{equation}
where $\vec{\nu_{ij}}$ is the relative momentum of particles $i$ and
$j$. In these equations $P_i$ is the pressure measured at the
position of the $i$-th particle.

Given the presence of a magnetic field, the magnetic contribution to
the pressure can be written as
\begin{equation}
P_{\rm mag}=\frac{B^{2}}{2\mu_{0}} ,
\end{equation}
with $\mu_{0} =4\pi\times 10^{-7}$ and $B$ is interpreted as the
{\em rms} value of the (tangled) field. If we assume that the
magnetic field is passively moved and squeezed by the gas we can
think of this magnetic pressure as being related to the baryonic
density in the same way as the thermal pressure but with a different
equation of state described by the parameter $\alpha$. Writing $P =
P_{\rm th} + P_{\rm mag}$ where $P_{\rm th}$ is the thermal pressure
(calculated using the standard SPH methods) we can accommodate both
forms in the standard SPH equations by simply adding an extra
effective pressure variable. Note, however that this approach has
severe limitations: there is no back-reaction of the field on the
gas; it cannot cope with large-scale magnetic structures which are
inherently anisotropic; and the polytropic model we use does not
rigorously conserve energy. One should not place too much literal
emphasis on the quantitative results we present, especially for high
densities.

\begin{figure}
\begin{center}
\leavevmode \epsfysize=7.5cm \epsfbox{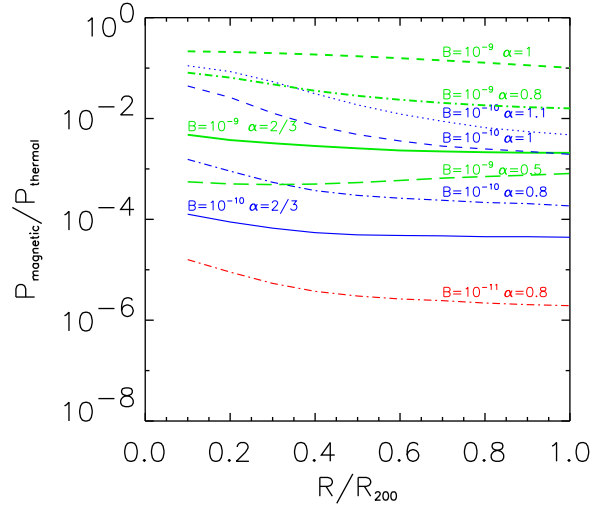}
\end{center}
\caption{Mean radial fractional pressure. Each line represents a
different run with the parameters labelled and displays the mean ratio of
the thermal to magnetic pressure in concentric spherical bins over all clusters.
\label{pressprof}}
\end{figure}

The magnetic field strength as a function of
overdensity at $z=0$ in our models is shown in
Figure~\ref{figmagprop}.
 We fixed the normalization factor $B_{0}$ and the power law index $\alpha$
 so that we span the range of observed field
strengths in halo cores as well as the values obtained by the more
complex model of Dolag \etal (2005) which is shown as the dashed
line. The top green shaded region represents the area in the $B-\rho$
plane where $B_{0} =10^{-9} G$ and $0.5 \le \alpha \le 1.$;
the lower blue region is relative to $B_{0} =10^{-10} G$ and $2./3. \le \alpha \le 1.1$
while the red continuous line is $B_{0} =10^{-11} G$ and $\alpha=0.8$.

For these values the highest normalisation models
challenge the latest observational constraints for the magnetic energy
density in clusters obtained from Faraday rotaton measures (Vogt \& Ensslin, 2006). Our models were chosen to span the full range of
possible magnetic field strengths, with the strongest fields larger
than observed (and so producing consequences larger than expected in the
observed Universe).

In uncollapsed regions the observed strength of the magnetic field
is currently only an upper limit. Several of our power-law models
exceed this value but as no structures collapse and we are not
analysing these volumes this has no affect on our results. Our
weakest field strength was chosen so as to fall within the range of
allowed values in the voids.

\begin{figure}
\begin{center}
\leavevmode \epsfysize=6.5cm \epsfbox{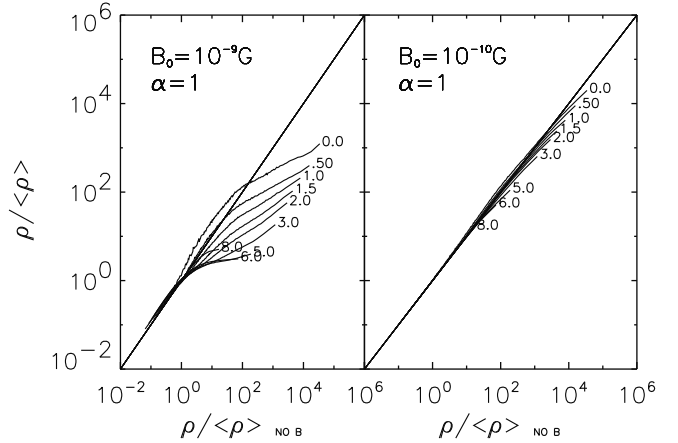}
\end{center}
\caption{A comparison at different redshifts of the densities of equivalent gas
particles from simulations without a magnetic field (noB) and one with
a magnetic field of strength $10^{-9} G$, $\alpha =1.$
(B0960, left panel) and a magnetic field of strength $10^{-10} G$,
$\alpha =1$ (B1060, right panel). The appropriate redshift is labelled at the
end of each profile.
\label{rhob9e6}}
\end{figure}

\onecolumn
\begin{figure}
\begin{center}
\leavevmode \epsfysize=7cm \epsfbox{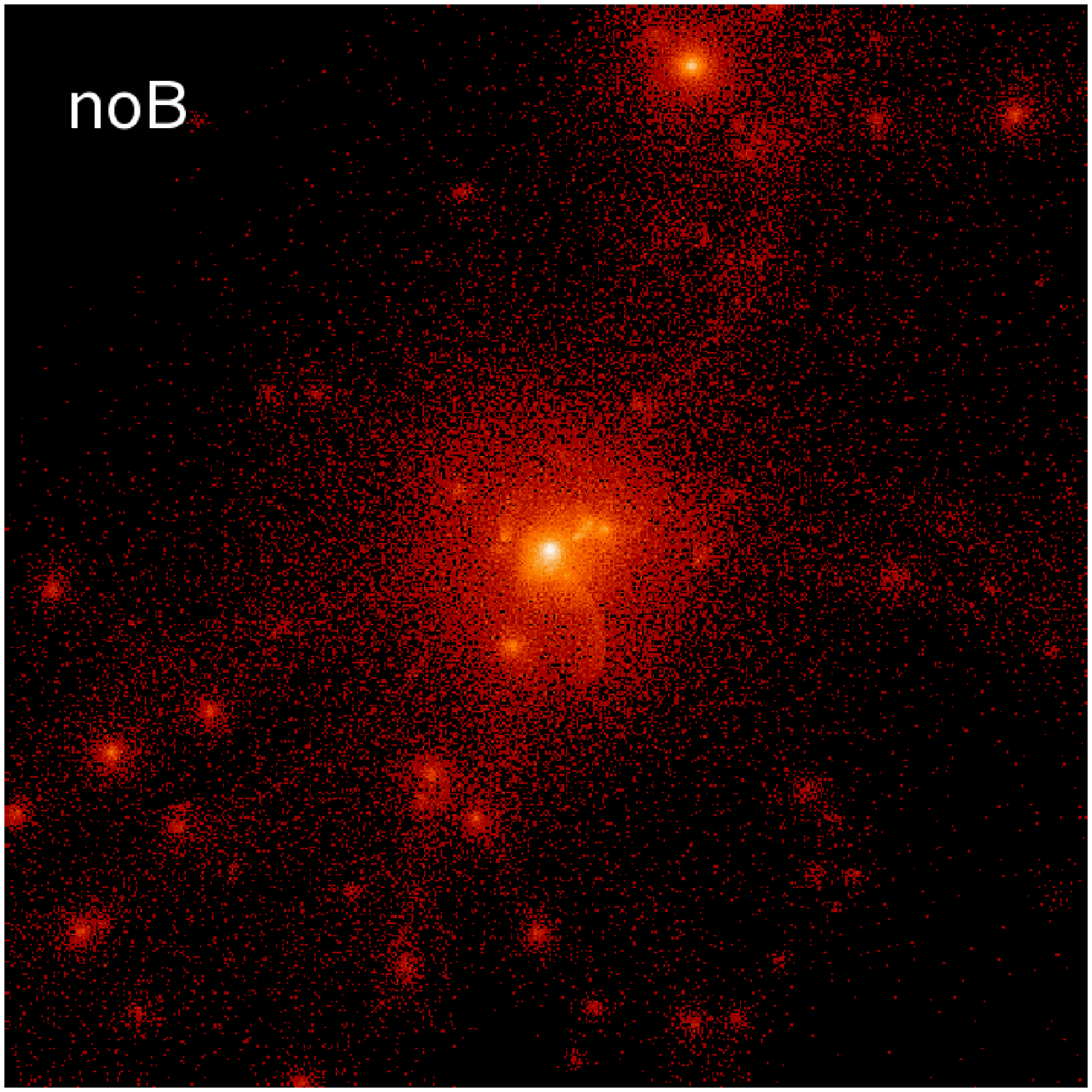}
\leavevmode \epsfysize=7cm \epsfbox{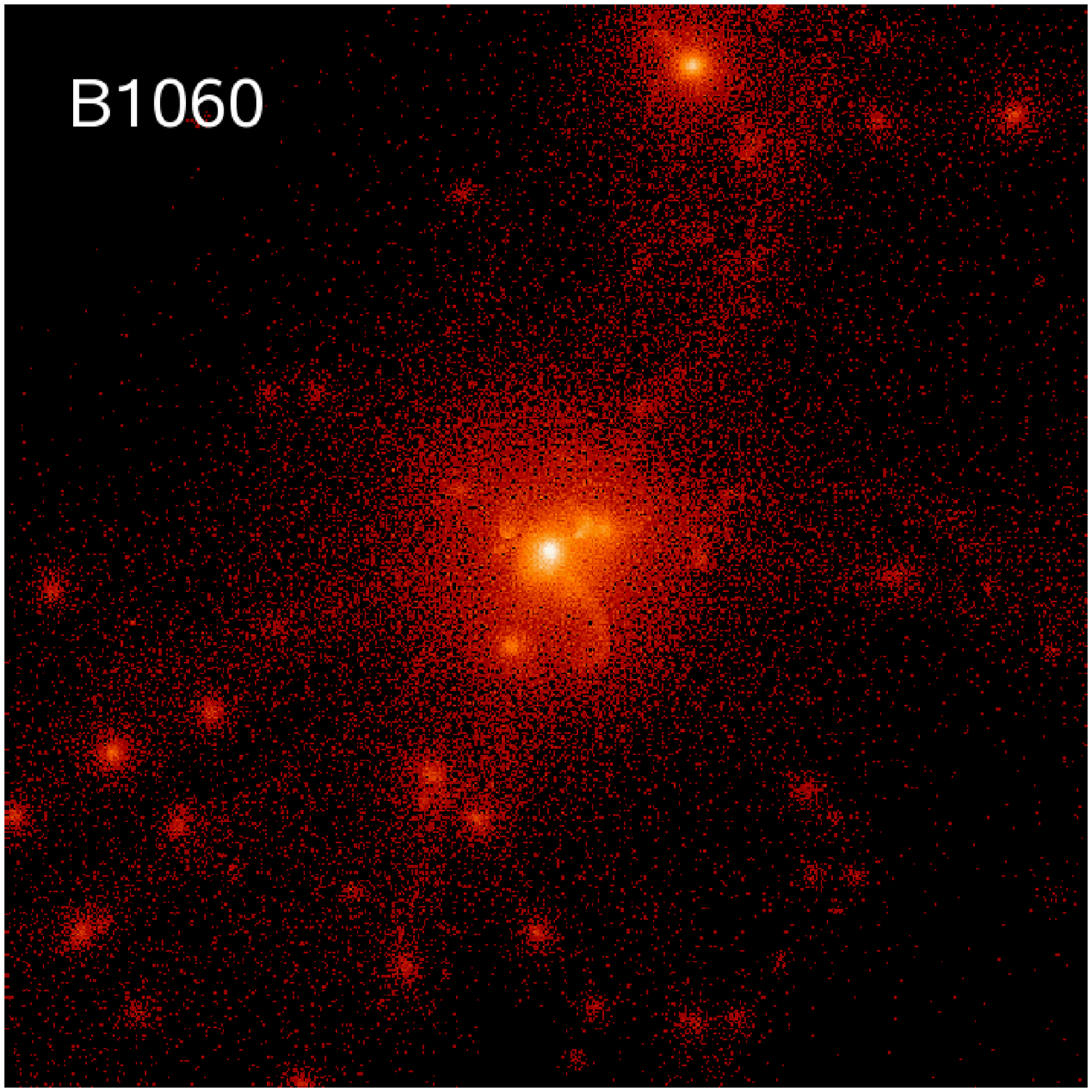}
\leavevmode \epsfysize=7cm \epsfbox{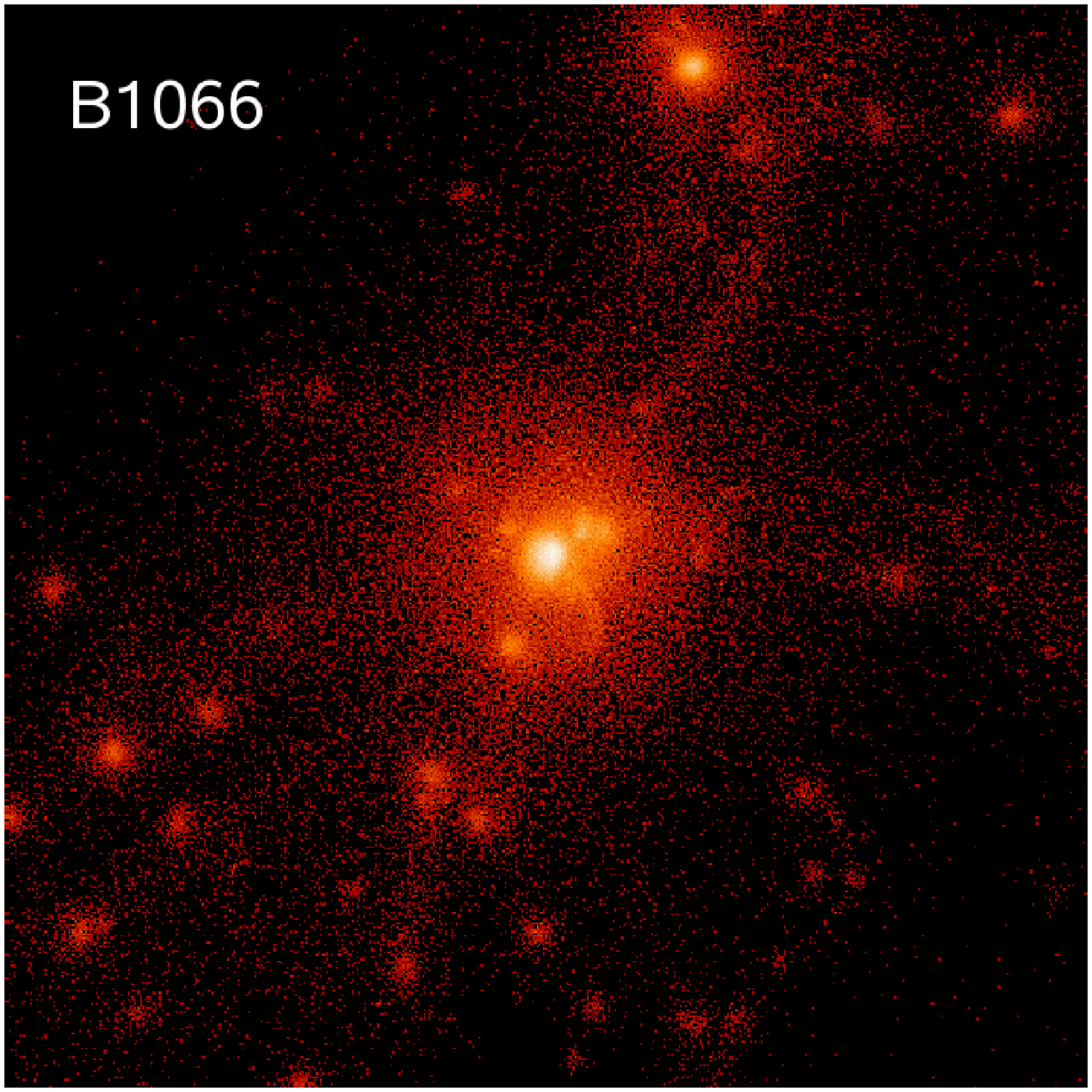}
\leavevmode \epsfysize=7cm \epsfbox{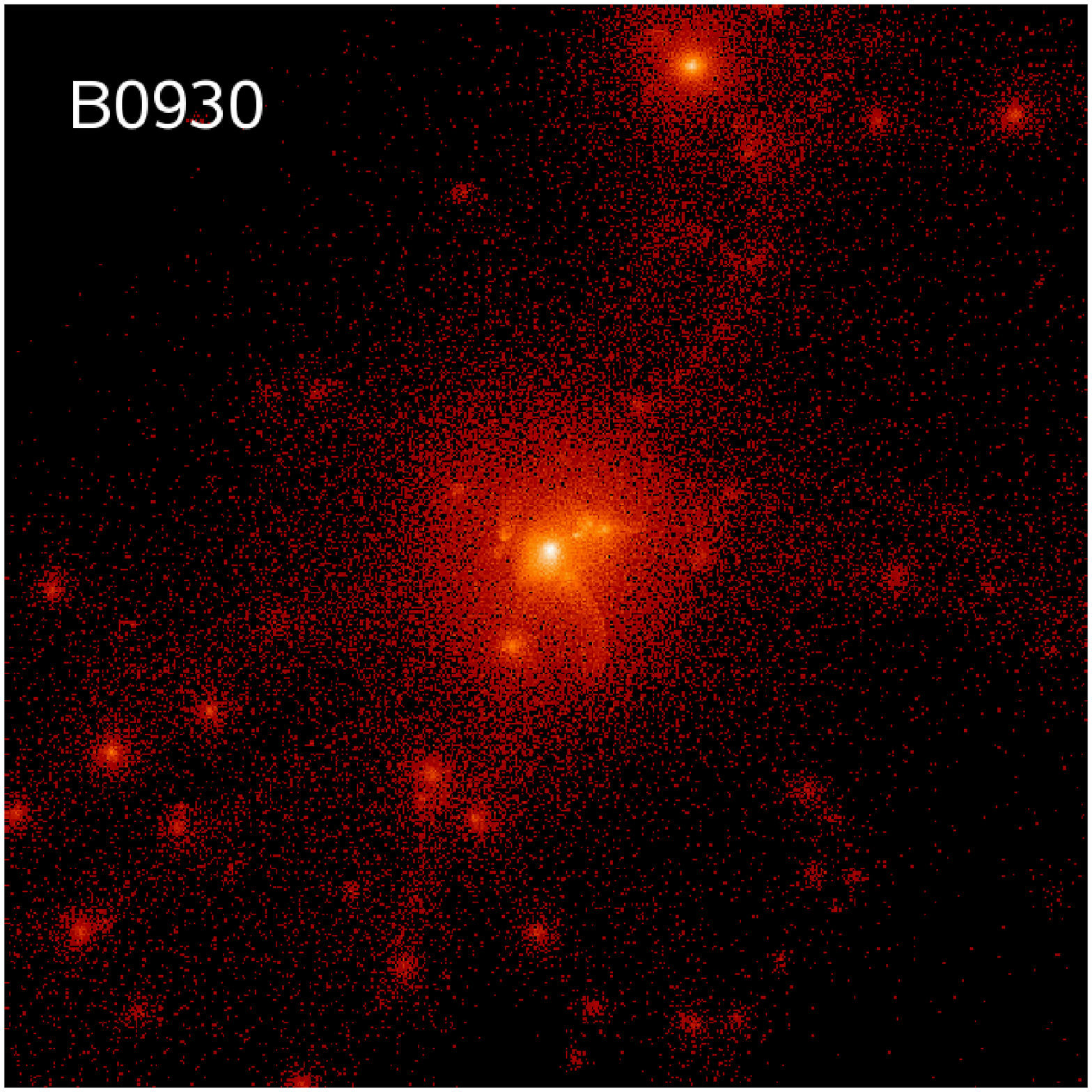}
\leavevmode \epsfysize=7cm \epsfbox{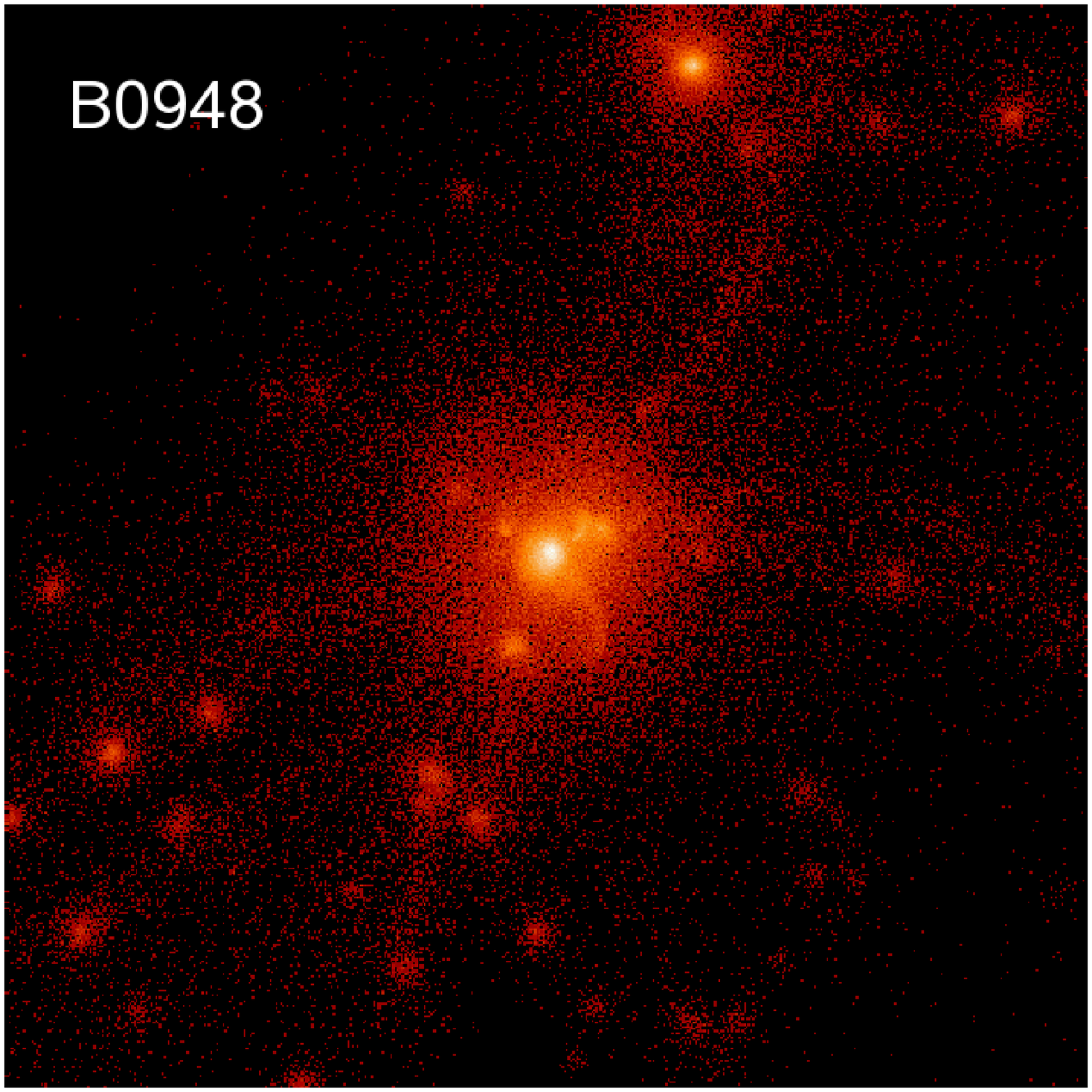}
\leavevmode \epsfysize=7cm \epsfbox{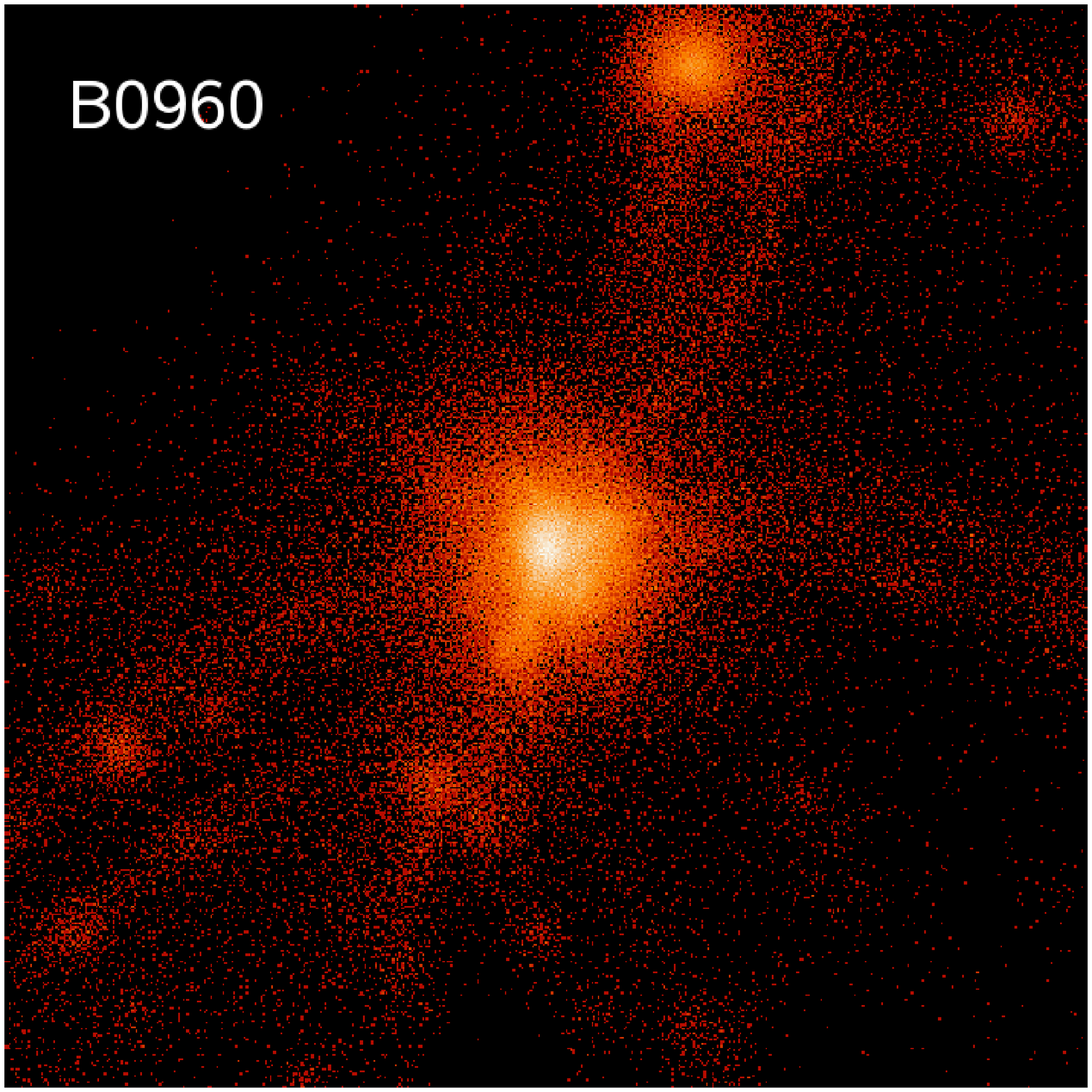}
\end{center}
\caption{Mass maps at $z=0$ for noB, B1060, B1066 (left, top to bottom); B0930, B0948 and B0960 (right, top to bottom). The box side is $10 h^{-1} {\rm Mpc}$ and the central object has a mass of $1.5\times10^{14}h^{-1} {\rm M}_{\odot}$.
\label{mapok}}
\end{figure}
\twocolumn

\noindent Once the amplitude and power-law index of the magnetic field have
been chosen there remains the choice of the redshift at which the
field is first imposed. If the magnetic field is turned on at late
times, after much of the structure has formed, then it has little
affect on the resultant objects. If, however, the field is imposed
before significant structures form (which in turn depends on the
resolution of the particular simulation being studied) then the
initial redshift makes little difference to the final distribution
of the matter.

Using trial simulations we established that for reasonable choices of initial field,
$z\ge5$ is necessary to have significant effect,
 but at higher redshifts the specific choice of z is not important. 
We therefore impose the magnetic field at $z=9$ for all subsequent runs.
\section{Redshift evolution}
We start our analysis with Figure ~\ref{pressprof} where we show the
ratio of $P_{\rm mag}$ versus $P_{\rm th}$ as a function of radii.
The values have been obtained averaging over all clusters and each
line is relative to a different run. The inclusion of a magnetic
field $B$ adds a non-thermal pressure $P_{\rm mag} \propto B^{2}$
that counteracts gravity. When $P_{\rm mag}$ is not negligible
compared to $P_{\rm th}$ the magnetic fields become dynamically
important in that the additional magnetic pressure supports the gas
against gravity and so for the same temperature the gas can reside
at a lower density. Also, the additional pressure can slow the
infall of the gas, retarding the evolution of structure. In reality
we found that $P_{\rm mag}$ is orders of magnitude smaller than
$P_{\rm th}$ in most cases and the ratio of gas magnetic to thermal
pressure generally falls with radius. However, even an
additional $1\%$ pressure support in the core region of collapsed
objects can have serious consequences, particularly for observables
such as the X-ray luminosity which depends on the square of the gas
density.

In Figure \ref{rhob9e6} we plot the rank-ordered density of all
the particles in the run with magnetic field compared to the rank
ordered density without a magnetic field at a range of different
redshifts. With the larger field (left panel) the high density
particles incur an appreciable drop in their density. At lower
field values (right panel) the effect of the magnetic field is
much reduced.

As can be seen for the high field redshift zero
case, at low redshifts and high field values particles near the
mean density actually end up denser than when no magnetic field
was present. This is due to the magnetic field pressure slowing
the infall and reducing the strength of the accretion shock,
resulting in a lower final entropy and consequently greater final
density. We find that $70\%$ of the particles with density ratio
$\rho_{\rm 0960}/\rho_{\rm noB} > 1$ have entropy ratio $s_{\rm
0960}/s_{\rm noB} < 1$, while the percentage increases to more
than $90\%$ if we consider particles with $\rho_{\rm
0960}/\rho_{\rm noB} > 10$ (where $\rho_{\rm 0960}$ ($s_{\rm
0960}$) and $\rho_{\rm noB}$ ($s_{\rm noB}$) are the densities
(entropies) for run $0960$ and noB).

In Figure \ref{mapok} we show a series of gas density maps (where brighter colours indicates
denser gas) of a box of side $10 h^{-1}{\rm Mpc}$ centered on
one of the most massive clusters ($1.5\times10^{14}h^{-1}{\rm M}_{\odot}$).
In this sequence of 6 panels the
field strength generally increases down the page with the precise
configuration of the magnetic field formulation that corresponds to
each panel given in Table~1. The presence of a strong magnetic field also results in a
smoother mass distribution, with small clumps
washed out and less frequent shocks.

\begin{figure}
\begin{center}
\leavevmode \epsfysize=7cm \epsfbox{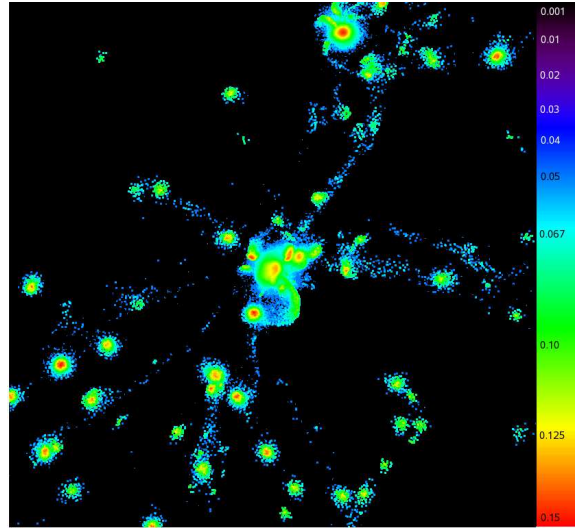}
\end{center}
\caption{Map of the ratio of magnetic to thermal pressure at $z =
0$ for B1060. The box is $10 h^{-1}{\rm Mpc}$ across, same region
as in figure \ref{mapok}. \label{mappmth2}}
\end{figure}

\begin{figure}
\begin{center}
\leavevmode \epsfysize=8.5cm \epsfbox{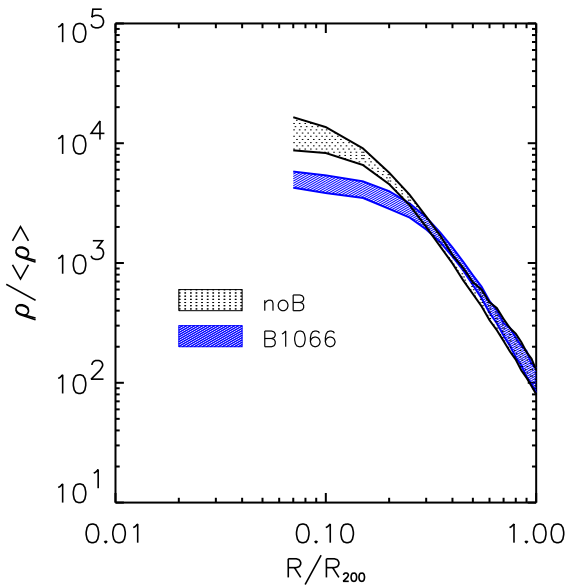}
\end{center}
\caption{Density profile at $z = 0$ for our clusters. The lines
represent the first and the third quartile, the dashed blue area is
for run B1066 and the white dotted one for the noB run. \label{denprof}}
\end{figure}

\begin{figure}
\begin{center}
\leavevmode \epsfysize=7.7cm \epsfbox{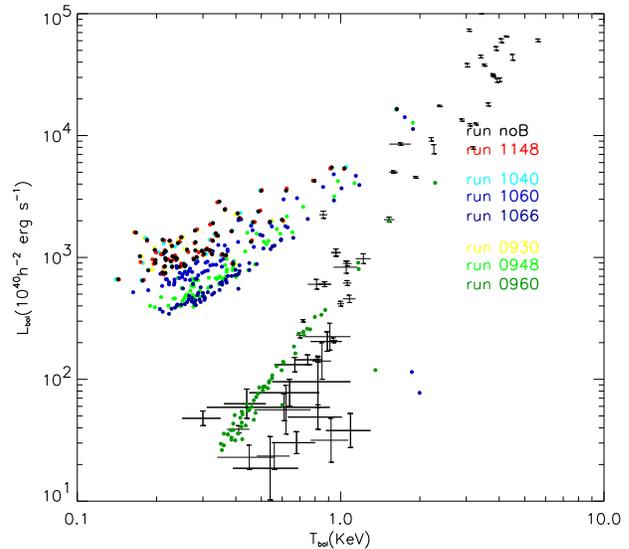}
\end{center}
\caption{The luminosity$-$temperature relation. The values plotted are the emission weighted temperature and the bolometric luminosity within $r_{500}$. Observations are from Ponman \etal 1996; Helsdon \etal 2000 and  Novicki \etal 2002.
\label{LTmag}}
\end{figure}

Spatially mapping the ratio of the magnetic to thermal pressure as
in Figure \ref{mappmth2} (which shows the same region as in Figure \ref{mapok})
reveals that only in small objects and cluster cores can the magnetic
pressure approach equipartition with the thermal pressure
($P_{\rm mag} \sim 0.1-0.3 P_{\rm th}$), provided that $B_{0}$ and
$\alpha$ are sufficiently high ($B_{0} \ge 10^{-10}$ and $\alpha \ge
1.$). This results in a shallower density profile
(Figure \ref{denprof}) and a lower entropy level in the inner
regions of collapsed objects, while the outskirts are not affected.
The mean radial gas density profile is displayed in Figure
\ref{denprof} which shows that field strengths not much higher than
those observed can significantly depress the central gas density.
Moreover we expect
 small structures to be affected more than big clusters because the former are
 characterised by a shallow gravitational
potential and so a relatively small additional pressure can be
sufficient to deter gas infall. This can be easily seen from
density maps (Figure \ref{mapok}) but, due to the small simulation
volume, we do not have a wide range in mass to properly
address any trend in mass from density profiles. Also,
unfortunately, our spatial resolution is not high enough to
properly investigate the regions where the magnetic field effects
should be more pronounced. 
In Figure \ref{LTmag} we show the
bolometric luminosity in erg $\rm s^{-1}$ versus the bolometric
emission weighted temperature in $\rm KeV$ for different runs. The
reduction in the central gas density lowers the bolometric X-ray
luminosity of a group-size object ($\rm{M} <10^{13}h^{-1}{\rm
M}_{\odot}$) by $40-70\%$ for field values of $B_{0}=10^{-10} G$
and $\alpha=1.1$ (Figure \ref{LTmag}). The reduction in luminosity
can be even more than one order of magnitude for the strongest
field we analysed.

\section{Observational consequences}

\begin{figure}
\begin{center}
\leavevmode \epsfysize=7.7cm \epsfbox{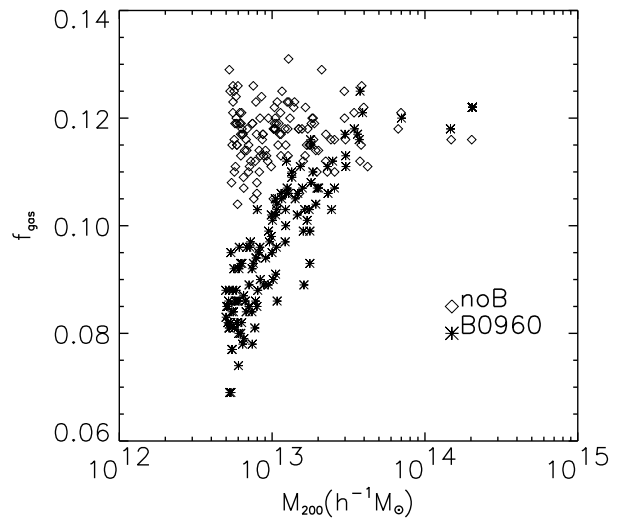}
\end{center}
\caption{Gas fraction within the virial radius as a function of the virial
mass at $z=0$ for a run with no magnetic field and with $B_{0}=10^{-9}G$ and $\alpha =1.$
\label{gasfrac}}
\end{figure}

When considering densities typical of collapsed objects at $z=0$ we
find that the average density profile of the haloes becomes shallower in the
core and the core entropy level increases when a magnetic field is
added to our simulations. On the other hand, the density profile in
the outskirts is not affected, even in the presence of a strong field
because it is mainly in the inner regions that the magnetic pressure
can reach equipartition with the thermal pressure (see
Figure \ref{pressprof}), while in the outskirts it is always
negligible.  This shallowing of the gas density profile results in a
more extended gas distribution and naturally leads to a reduction in
the baryon fraction within our objects in the presence of a magnetic
field (Figure ~\ref{gasfrac}) (see Ettori \etal (2006) for a wider
discussion of this topic). For a reasonably high field strength and
slope (run B0960) the baryon fraction is reduced to 60\% of the cosmic
value, with the effect becoming more dramatic as the halo mass
decreases.

\begin{figure}
\begin{center}
\leavevmode \epsfysize=8cm \epsfbox{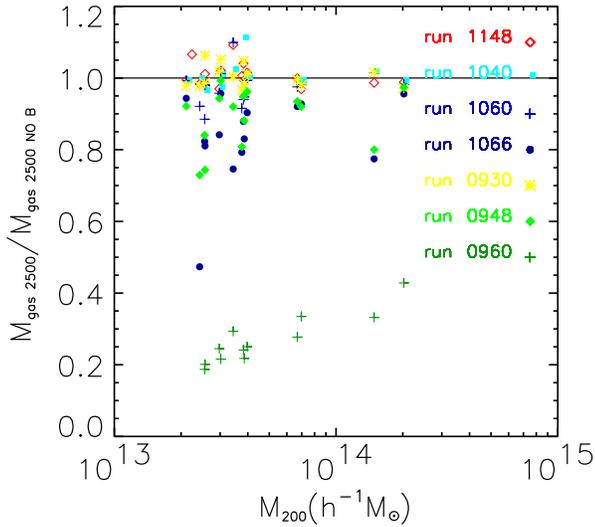}
\end{center}
\caption{Relative gas mass within $r_{2500}$ (the radius enclosing an
overdensity of 2500) in runs with a magnetic field versus the total virial
mass in the run without a magnetic field. The different symbols mark different runs.
\label{denprof2500}}
\end{figure}

Figure~\ref{denprof2500} displays the ratio of the
gas mass within $r_{2500}$ (the radius that encloses an overdensity of $2500$) for the models with various magnetic field
strengths compared to the gas fraction obtained in the run without a
magnetic field.
For regions with overdensities exceeding 2500 times the mean (which
corresponds to the halo core region which is typically well observed in
X-rays) the baryonic mass fraction can be even more dramatically reduced,
although the effect is only particularly pronounced for higher
field strengths.

\section{Summary \& Conclusions}

We have presented a simple approximate scheme that implements an isotropic
magnetic field within standard SPH. This
method allows various field strength and amplification schemes to be
quickly and efficiently tested. Using our approach we have modelled
the full range of observed field strengths and shown that the stronger
fields can have a significant affect on the halo core regions.

It is important to note that this work could equally well be applied
to any physical process that supplements the usual gas pressure
support with additional pressure which has a power-law relationship
to the density. This could include local turbulence (Dolag \etal
2006), the injection of hot gas from supernovae (Silich,
Tenorio-Tagle \& Anorve-Zeferino 2005), radio bubbles (Ensslin 2003;
Soker \& Pizzolato 2005; Fabian \etal 2006), or the activity of
galactic nuclei (Sijacki \& Springel 2006).  As would perhaps have
been expected smaller haloes are more significantly altered by these
processes, particularly when there is a strong density dependence.

Any physical process that effects the central baryonic density within
dark matter haloes is astrophysically interesting because such a
process will naturally alter star formation rates, feeding of any
central black hole and the global X-ray emission from the object (to
name but a few). Even a small lowering of the central density can have
a dramatic effect because many of these associated processes operate
on a high power of the local density (for instance, the X-ray emission,
largely due to thermal bremsstrahlung, is related to the density squared).

If it indeed turns out that large scale astrophysical magnetic fields
are similar in strength to those suggested by Dolag \etal (2001) then in
the main they will have little effect, particularly if they remain
untangled and so amplify with the density to a low power. If the magnetic
field in collapsed regions becomes tangled however, and so grows with
a steeper power law dependence on density then their affects could be
significant, particularly within the very central regions of small haloes.

\section*{Acknowledgements}
We thank the referee for his thoughtful and detailed report which
has allowed us to substantially improve the paper. LG acknowledges a University of Nottingham research studentship. The simulation work was performed on the Nottingham HPC facility.

\end{document}